\documentclass[reviewcopy,final,5p,times,twocolumn]{elsarticle}
\usepackage[reviewcopy]{adndt}
\usepackage{longtable}
\usepackage{booktabs}
\usepackage{bm}

\usepackage{amsmath}
\usepackage{color}
\usepackage{soul}
\usepackage{ulem}
\usepackage{amssymb}
\usepackage{graphicx}
\biboptions{square,sort&compress}
\bibpunct[]{[}{]}{,}{n}{}{;}
\newcommand{\journaltitle}{Atomic Data and Nuclear Data Tables}
\newcommand{\journalhome}{\texttt{http://www.elsevier.com/locate/jnlabr/yadndt}}
\newcommand{\journalmail}{\texttt{adndt@elsevier.com}}
\newcommand{\cmd}[1]{\texttt{$\backslash${#1}}}
\newcommand{\templatename}{\texttt{tmpadndt.tex}}
\newcommand{\adndtstyle}{\texttt{adndt}}
\newcommand{\adndtbst}{\texttt{adndt.bst}}
\newcommand{\adndtguide}{\texttt{../doc/ADNDdoc.pdf}}

\newcommand{\dprime}{{\prime\prime}}
\newcommand{\tprime}{{\prime\prime\prime}}
\newcommand{\fprime}{{\prime\prime\prime\prime}}
\newcommand{\te}[1]{\textrm{#1}}

\newcommand{\vdag}{(v)^\dagger}
\newcommand\aastex{AAS\TeX}
\newcommand\latex{La\TeX}

\journal{ADNDT, prepared by using elsarticle.cls}
\def\kw#1{{\color{blue} #1}}
\begin{document}

\begin{frontmatter}

\journal{Atomic Data and Nuclear Data Tables}


\title{Extended Calculations with Spectroscopic Accuracy: Energy Levels and Radiative Rates for O-like Ions between Ar XI and Cr XVII}


\author[HB,IMP]{C. X. Song}
\author[IMP]{C. Y. Zhang}
\author[HB,IMP]{K. Wang\corref{corl}}
\ead{wang$_{-}$kai10@fudan.edu.cn}
\author[ULB,IMP]{R. Si\corref{corl}}
\ead{rsi13@fudan.edu.cn }
\author[ULB]{M. Godefroid}
\author[Malmo]{P. J{\"o}nsson}
\author[HB]{D. Wei\corref{corl}}
\ead{dangwei@hbu.edu.cn}
\author[HB]{X. H. Zhao}
\author[BJ]{J. Yan}
\author[IMP]{C. Y. Chen}

\cortext[corl]{Corresponding Author}

\address[HB]{Hebei Key Lab of Optic-electronic Information and Materials, The College of Physics Science and Technology, Hebei University, Baoding 071002, China;}
\address[IMP]{Shanghai EBIT Lab, Key Laboratory of Nuclear Physics and Ion-beam Application, Institute of Modern Physics, Department of Nuclear Science and Technology, Fudan University, Shanghai 200433, China; }
\address[ULB]{Spectroscopy, Quantum Chemistry and Atmospheric Remote Sensing (SQUARES), CP160/09, Universit\'{e} libre de Bruxelles, Av. F.D. Roosevelt 50, 1050 Brussels, Belgium;}
\address[Malmo]{Department of Materials Science and Applied Mathematics, Malm{\"o} University, SE-20506, Malm{\"o}, Sweden;}
\address[BJ]{Institute of Applied Physics and Computational Mathematics, Beijing 100088, China;}

\begin{abstract}
Using the multiconfiguration Dirac-Hartree-Fock and the relativistic configuration interaction methods,
a consistent set of transition energies and radiative transition data for the main states of the $2s^2 2p^4$, $2s 2p^5$, $2p^6$, $2s^2 2p^3 3s$, $2s^2 2p^3 3p$, $2s^2 2p^3 3d$, $2s 2p^4 3s$, $2s 2p^4 3p$, and $2s 2p^4 3d$ configurations in  O-like Ions between Ar XI  ($Z = 18$) and Cr XVII ($Z = 24$) is provided. 
Our data set is compared with the NIST compiled values and previous calculations. The data are accurate enough for identification and deblending of new emission lines from hot astrophysical and laboratory plasmas. The amount of data of high accuracy is significantly increased for the $n = 3$ states of several O-like ions, where experimental data are very scarce.  
\end{abstract}

\end{frontmatter}




\newpage{\onecolumn\tableofcontents \listofDtables \listofDfigures \vskip4pc}

\twocolumn
\section{Introduction}
Oxygen-like ions with nuclear charge numbers from eighteen to twenty four have long been observed in hot astrophysical and laboratory plasmas~\citep{Behring.1972.V175.p493,McKenzie.1982.V254.p309,Feldman.1998.V503.p467,Feldman.2004.V607.p1039,Traebert.2018.V865.p148}.          
For example, spectra of Ar~XI, Ca~XIII, Ti~XV, and  Cr~XVII have been observed in  the solar corona and in flares~\citep{Feldman.2000.V544.p508,Curdt.2004.V427.p1045,DelZanna.2012.V537.p38,DelZanna.2018.V852.p52}. 
Though strong lines from these ions of astrophysical abundant elements (Ar, Ca, K, Ti, and Cr) are well known, a large fraction of the lines from these ions are unclassified due to the lack of accurate atomic data~\citep{Beiersdorfer.2014.V788.p25,Traebert.2014.V215.p6,Traebert.2014.V211.p14,Beiersdorfer.2018.V854.p114,Traebert.2018.V865.p148}. 
To remedy this situation,  we have reported highly accurate excitation energies and radiative transition rates for L-shell and M-shell atomic ions of the above elements~\citep{Wang.2014.V215.p26,Wang.2015.V218.p16,Wang.2016.V223.p3,Wang.2016.V226.p14,Wang.2017.V229.p37,Wang.2018.V235.p27,Wang.2018.V239.p30,Wang.2020.V246.p1,Si.2016.V227.p16,Si.2018.V239.p3}. 
This work extends our effort to accurate atomic data of O-like from Ar~XI to Cr~XVII.

Using different methods, excitation energies and/or radiative decay rates of the $(1s^2) 2s^2 2p^4$, $2s 2p^5$, and $2p^6$ configurations for O-like from Ar~XI to Cr~XVII have been provided~\citep{Cheng.1979.V24.p111,Fischer.1983.V28.p3169,Baluja.1988.V21.p1455,Gaigalas.1994.V49.p135,Safronova.1999.V60.p36,Zhang.2002.V82.p357,Landi.2005.V434.p365,Gu.2005.V89.p267,Rynkun.2013.V557.p136}. 
Because of wide applications in modeling and diagnosing different kinds of plasmas, and in analyzing new observations of different astrophysical sources,  energy structures and radiative transition parameters for higher-lying states of the $n \geq 3$ configurations are eagerly needed~\citep{Feldman.2000.V544.p508,Feldman.2004.V607.p1039,Shestov.2008.V34.p33}. {\sc autostructure} calculations for the lowest 86 levels of the $2s^2 2p^4$, $2s 2p^5$, $2p^6$, $2s^2 2p^3 3s$, $2^2 2p^3 3p$, and $2s^2 2p^3 3d$ configurations in Ar~XI and Ca~XIII were provided by~\cite{Landi.2006.V92.p305,Landi.2005.V444.p305}. 
\citet{Bogdanovich.2008.V94.p623} reported excitation energies and radiative transition data for the lowest 200 levels of the $n \leq 3$ configurations in Cr~XVII  using a configuration interaction method. In their calculations, the relativistic effects were taken into account within the traditional Breit–Pauli approximation. 

Among the previous theoretical studies, the results for the $n=2$ levels provided by \citet{Gu.2005.V89.p267} and \citet{Rynkun.2013.V557.p136} are, so far, the most accurate.
By contrast, previous calculations ~\citep{Landi.2006.V92.p305,Landi.2005.V444.p305,Bogdanovich.2008.V94.p623} involving the $n \leq 3$ levels of Ar~XI, Ca~XIII, and Cr~XVII are indeed not accurate enough due to restricted configuration interaction effects included in their works.  
Excitation energies of the calculations by~\citet{Landi.2006.V92.p305} for Ar~XI, by~\citet{Landi.2005.V444.p305} for Ca~XIII, and by~\citet{Bogdanovich.2008.V94.p623} for Cr~XVII  differ from the corresponding experimental values in the Atomic Spectra Database (ASD) of the National Institute of Standards and Technology (NIST)~\citep{Kramida.2018.V.p} by up to 35000 cm$^{-1}$, 35000 cm$^{-1}$ and 20000 cm$^{-1}$, respectively. 
Therefore, systematic calculations of high accuracy involving the $n \geq 3$ states are greatly desired. 
Databases such as CHIANTI \citep{Dere.2019.V241.p22,Dere.1997.V125.p149} also demand complete and consistent data sets of high accuracy, with the view of offering the astrophysical community tools and data to carry out accurate plasma diagnostics.

Recently, using the multireference M\o ller-Plesset (MR-MP) perturbation method~\citep{Santana.2018.V234.p13,Santana.2019.V245.p9,Santana.2020.V247.p52}, \citet{Santana.2018.V238.p34} reported calculated excitation energies for L-shell ions of Argon. In their calculations, they adopted  large configuration state function expansions and reported calculated excitation energies with high accuracy, including the $n \leq 3$ levels in Ar XI.  Their data can be used to identify observed spectral lines.

The present work provides a consistent data sets of energy structures and radiative transition data with high accuracy for  O-like ions in the range of nuclear charges $18 \le Z \le 24$.
Using the multiconfiguration Dirac-Hartree-Fock (MCDHF) and the relativistic configuration interaction (RCI) methods implemented in the GRASP2K code~\citep{Jonsson.2013.V184.p2197}, excitation energies and lifetimes for the main $n \leq 3$ in  Ar XI  (K XII, Ca XIII, Sc XIV, Ti XV, V~XVI, Cr XVII), and multipole transition rates (electric dipole (E1), magnetic dipole (M1), and  electric quadrupole (E2)) among these states are calculated.  Compared with previous studies of O-like ions, our calculations result in a significant extension of accurate energy and transition data for higher-lying states of the $n = 3$ configurations, which will greatly improve the assessment of blending for diagnostic lines of interest, and aid the analysis of new spectra from hot astrophysical and laboratory plasmas.

\section{Theory and Calculations}
The MCDHF method in the GRASP2K code~\citep{Jonsson.2013.V184.p2197,Jonsson.2007.V177.p597} is reviewed  by~\citet{FroeseFischer.2016.V49.p182004}. This method is also described in our recent papers~\citep{Wang.2017.V117-118.p1,Wang.2017.V117-118.p174,Wang.2018.V123.p114,Wang.2019.V236.p106586,Wang.2019.V237.p106640,Wang.2017.V119.p189301,Li.2020.V133134.p101339,Li.2019.V126.p158,Zhao.2018.V119.p314,Song.2020.V247.p70,Chen.2019.V129-130.p101278,Chen.2018.V206.p213,Chen.2017.V113.p258,Chen.2017.V114.p61}. For this reason, in the sections below, only the computational strategies are described.

In our MCDHF calculations, the multireference (MR) sets for even and odd parities include

even configurations:
$2s^{2}2p^{4}$, $2p^{6}$, $2s^{2}2p^{3}3p$, $2s2p^{4}3s$, $2s2p^{4}3d$, $2p^{5}3p$, $2s^{2}2p^{3}4p$, $2s^{2}2p^{3}4f$;


odd  configurations:
$2s2p^{5}$, $2s^{2}2p^{3}3s$, $2s^{2}2p^{3}3d$, $2s2p^{4}3p$, $2p^{5}3s$, $2p^{5}3d$, $2s^{2}2p^{3}4s$, $2s^{2}2p^{3}4d$.

We start the calculation without any excitation from the MR configurations, which is usually referred to as the Dirac-Fock (DF) calculation. 
Based on the active space (AS) approach~\cite{Sturesson.2007.V177.p539} for the generation of the configuration state function (CSF) expansions, separate MCDF calculations are done for the even and odd parity states. 
Subsequently, the CSFs expansions are obtained through the  single and double (SD)  substitutions from subshells of the reference configurations up to a $ nl $ orbital, with $n \leq 8$ and $l \leq 5$. 
To reduce the number of CSFs, the $1s^2$ core is closed during the relativistic self-consistent field (RSCF) calculations, but is opened during the following RCI calculations, where the Breit interaction and the leading  QED effects, i.e., vacuum polarization and self- energy, are included in the Hamiltonian. 
The number of CSFs in the final even and odd state expansions are
approximately 1~796~000  and 3~938~000, respectively, distributed over the different $J$ symmetries.

By  using the $jj$-$LSJ$ transformation approach~\citep{Gaigalas.2017.V5.p6, Gaigalas.2004.V157.p239}, the  $jj$-coupled CSFs are transformed into $LSJ$-coupled CSFs, from which the $LSJ$ labels used by experimentalists are obtained.

\section{Results and discussions}

\subsection{Excitation energies}~\label{Sec:en}
Our MCDHF/RCI excitation energies $E_{\rm MCDHF/RCI}$ (cm$^{-1}$), together with MCDHF/RCI radiative lifetimes $\tau_{\rm MCDHF/RCI}^l$ (in~s) in the length  form and $\tau_{\rm MCDHF/RCI}^v$ (in s)  in the velocity form, 
are reported in Table~\ref{tab.lev.all} for the lowest 156 (179, 184, 196, 200, 200, 200) states of the $2s^2 2p^4$, $2s 2p^5$, $2p^6$, $2s^2 2p^3 3s$, $2s^2 2p^3 3p$, $2s^2 2p^3 3d$, $2s 2p^4 3s$, $2s 2p^4 3p$, and $2s 2p^4 3d$ configurations in  Ar XI  (K XII, Ca XIII, Sc XIV, Ti XV, V~XVI, Cr XVII). All these states are below the first $2s^2 2p^3 4s$ level. 
Observed values $E_{\rm NIST}$   (cm$^{-1}$) from the NIST ASD \citep{Kramida.2018.V.p} are also listed in this table. 
As many levels are strongly mixed, their configuration label is not unique.
Here, the parity, $J$ value and energy  are used to match the levels in the NIST ASD. 
The levels in which our $LSJ$-coupled labels are different from those in the NIST ASD are listed in Table~\ref{table.diff} for reference.

Experimental and theoretical energy data are available for all $n = 2 $ levels  along the isoelectronic sequence from Ar~XI to Cr~XVII. Computed excitation energies of the $n=2$ levels from the present MCDHF/RCI calculations, from the MCDHF/RCI calculations by~\citet{Rynkun.2013.V557.p136} [hereafter referred to as MCDHF/RCI2], from the many-body perturbation theory (MBPT) calculations by~\citet{Gu.2005.V89.p267} [MBPT],  from the SUPERSTRUCTURE calculations by~\citet{Landi.2005.V434.p365} [SS],  are compared with experimental values from the NIST ASD~\citep{Kramida.2018.V.p} in Table~\ref{table.n2}. Compared with the SS calculations, the present MCDHF/RCI calculations, as well as the MCDHF/RCI2 and MBPT calculations show better agreement with the NIST experimental values. The average absolute difference with the standard deviation~\citep{Wang.2017.V229.p37} between computed excitation energies and the NIST values for all $n = 2$ levels from Ar~XI to Cr~XIV are $4 \pm 197$ cm$^{-1}$ for MCDHF/RCI, $252 \pm 242$ cm$^{-1}$ for MCDHF/RCI2, and $-421 \pm  940$ cm$^{-1}$ for MBPT. The corresponding values for SS are $9154 \pm 10208$ cm$^{-1}$.

Among the seven O-like ions considered here, theoretical energy data of the lowest 200 levels up to the $n=3$ configurations for Cr XVII were provided by  \citet{Bogdanovich.2008.V94.p623} using a configuration interaction method. In our previous work, we also performed the calculations for the lowest 200 levels employing two state-of-the-art methods, MCDHF/RCI and MBPT. In what follows we will further access the accuracy of the present MCDHF/RCI excitation energies in Cr XVII, by comparing available results. In Table~\ref{table.Cr}, we present the MCDHF/RCI excitation energies of the lowest 200 levels for Cr XVII.  Observed energies from the NIST ASD 
and computed values from different sources (MCDHF/RCI3: the MCDHF/RCI results from \citet{Wang.2017.V229.p37}; MBPT2: the MBPT results from \citet{Wang.2017.V229.p37}; CI: theoretical  results from \citet{Bogdanovich.2008.V94.p623}) are also included. As can be seen in Table~\ref{table.Cr}, good agreement is obtained among the MCDHF/RCI, MCDHF/RCI3  and MBPT calculations for the lowest 200 levels. The average absolute difference with the standard deviation between computed excitation energies and the MCDHF/RCI values for all the $n \leq 3$ levels are $821
\pm 274$ cm$^{-1}$ for MCDHF/RCI3, and $202
\pm  497$ cm$^{-1}$ for MBPT2. This corresponds to an average relative difference of $0.014\% \pm
0.009\%$ for MCDHF/RCI3  and $0.002\% \pm
0.018\%$ for MBPT2, respectively, where the standard deviations are indicated after the values. The CI excitation energies from \citet{Bogdanovich.2008.V94.p623} are generally higher than our MCDHF/RCI values by ten to twenty of thousand cm$^{-1}$ for the $n=3$ levels.   The average absolute difference with the standard deviation from the present MCDHF/RCI values are $15 267 \pm
4 016$ cm$^{-1}$ for the CI calculations.

Compiled values from the NIST ASD are also questionable for a few $n = 3$ states of Cr  \uppercase\expandafter{\romannumeral17}, for which the deviations from the present results  are about or larger than 10~000 cm$^{-1}$.  For example, the values 6 074 000 ~cm$^{-1}$ for  the level  $2s^{2}\,2p^{3}(^{2}_{3}D)~^{2}D\,3d~^{3}D_{3}^{\circ}$ with the key ($\#68$),   
6 131 000  ~cm$^{-1}$ for the level  $\#80 / 2s^{2}\,2p^{3}(^{2}_{1}P)~^{2}P\,3d~^{3}P_{2}^{\circ}$, and 6 164 800 ~cm$^{-1}$ for the level $\#87 /  2s^{2}\,2p^{3}(^{2}_{1}P)~^{2}P\,3d~^{3}D_{3}^{\circ}$  do not have obvious counterparts in the present MCDHF/RCI calculations, as well as in the previous MCDHF/RCI3 and MBPT2 calculations.

\begin{figure}
	\includegraphics[width=10cm]{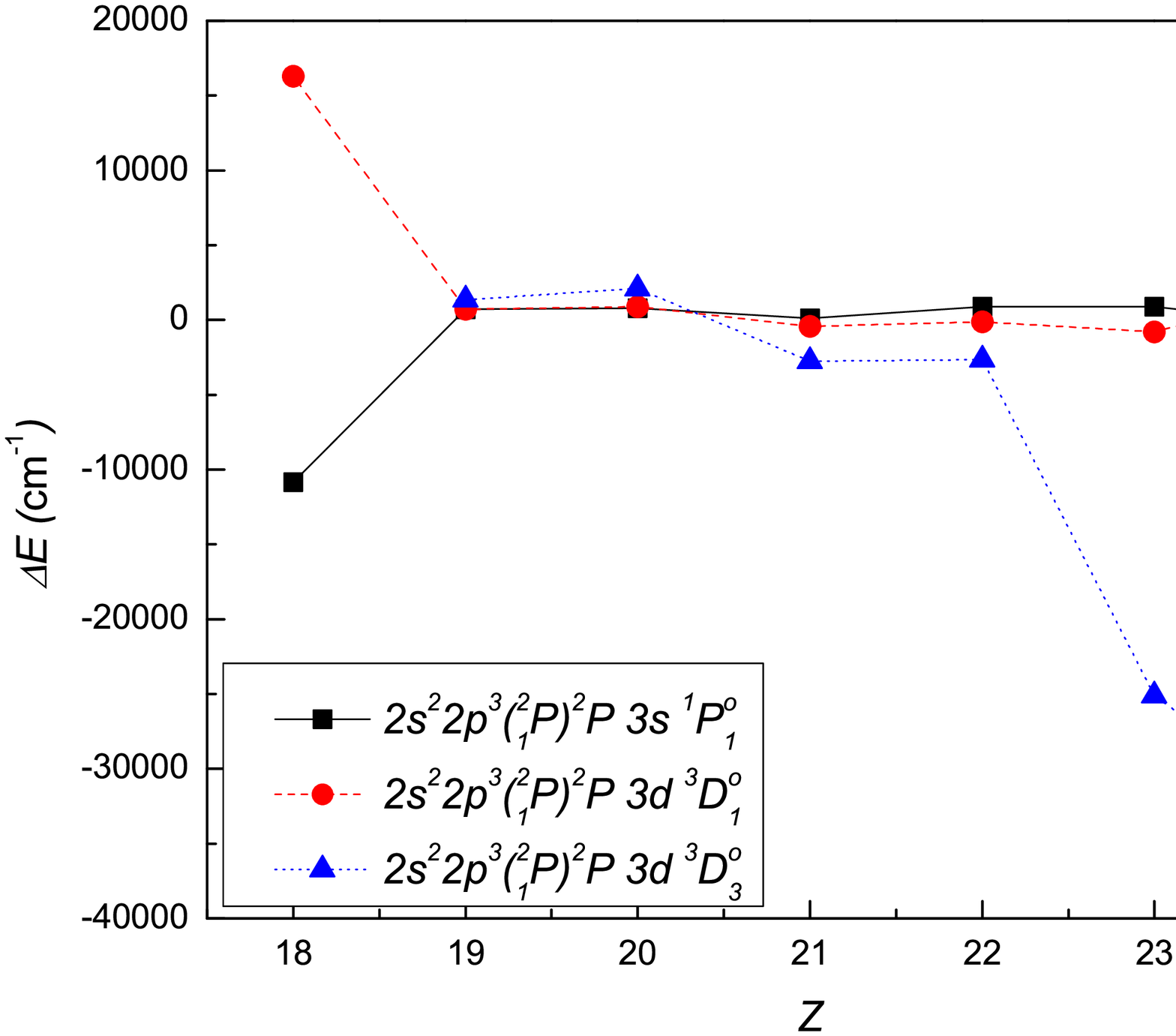}
	\caption{\label{fig2}Energy deviations $\Delta E = E_{\rm NIST}-E_{\rm MCDHF/RCI}$ in cm$^{-1}$ as a function of $Z$ for three levels ($2s^{2}\,2p^{3}(^{2}_{1}P)~^{2}P\,3s~^{1}P_{1}^{\circ}$, $2s^{2}\,2p^{3}(^{2}_{1}P)~^{2}P\,3d~^{3}D_{1}^{\circ}$, and $2s^{2}\,2p^{3}(^{2}_{1}P)~^{2}P\,3d~^{3}D_{3}^{\circ}$).  \label{fig.lev.nistlargedifferences}}
\end{figure}

The NIST compiled values differ from our MCDHF results by over 3000 cm$^{-1}$ (about 0.1\%) for many states of seven O-like ions considered in our work, and all these values have been listed in Table~\ref{table.nist}. As shown in our studies~\citep{Si.2016.V227.p16,Wang.2014.V215.p26,Wang.2015.V218.p16,Wang.2016.V223.p3,Wang.2016.V226.p14,Wang.2017.V229.p37,Wang.2018.V234.p40,Wang.2020.V246.p1,Song.2020.V247.p70}, the plot of differences between calculated  and measured excitation energies along the isoelectronic sequence is an useful tool to identify the possible of measured values with large uncertainty.  As an example, we show the differences (in cm$^{-1}$) between the present MCDHF/RCI excitation energies and the NIST values for three levels  ($2s^{2}\,2p^{3}(^{2}_{1}P)~^{2}P\,3s~^{1}P_{1}^{\circ}$, $2s^{2}\,2p^{3}(^{2}_{1}P)~^{2}P\,3d~^{3}D_{1}^{\circ}$, and $2s^{2}\,2p^{3}(^{2}_{1}P)~^{2}P\,3d~^{3}D_{3}^{\circ}$) as a function of $Z$ with $Z=18-24$ in Figure~\ref{fig.lev.nistlargedifferences}. The differences for the NIST values from the present MCDHF results are about -10~000 cm$^{-1}$ for the level $2s^{2}\,2p^{3}(^{2}_{1}P)~^{2}P\,3s~^{1}P_{1}^{\circ}$ in Ar XI,  about 15~000 cm$^{-1}$ for the level $2s^{2}\,2p^{3}(^{2}_{1}P)~^{2}P\,3d~^{3}D_{1}^{\circ}$ in Ar XI, and -35 000 cm$^{-1}$ -- -25 000 cm$^{-1}$ for the level $2s^{2}\,2p^{3}(^{2}_{1}P)~^{2}P\,3d~^{3}D_{3}^{\circ}$ in V~XVI  and Cr~XVII. Whereas good agreement is obtained for the same levels in the other ions along the isoelectronic sequence.  Since in the present MCDHF/RCI calculations the same computational strategies are used for each ion, the accuracy of our calculated excitation energies  should be consistent and systematic for the same level along the sequence.

We also showed the deviations of the NIST excitation energies to the  MCDHF/RCI3 results for two levels $2s^{2}\,2p^{3}(^{4}_{3}S)~^{4}S\,3d~^{3}D_{3}^{\circ}$ and $2s^{2}\,2p^{3}(^{2}_{3}D)~^{2}D\,3d~^{3}D_{3}^{\circ}$ of O-like ions  as a function of the  nuclear charge $Z$ with $Z=24 -30$ in Figure~1 (a) of our previous work~\citep{Wang.2017.V229.p37}. The deviations between the MBPT2 and MCDHF/RCI3 values for the same levels  along the sequence were shown in Figure~1 (b) of our previous work. A few of the NIST compiled values, including excitation energies for the two levels  $2s^{2}\,2p^{3}(^{4}_{3}S)~^{4}S\,3d~^{3}D_{3}^{\circ}$ and $2s^{2}\,2p^{3}(^{2}_{3}D)~^{2}D\,3d~^{3}D_{3}^{\circ}$ of Cr XVII listed in Table~\ref{table.nist}, depart from the MCDHF/RCI3 data by over than 4~000 cm$^{-1}$, while good agreement is obtained between the computed MCDHF/RCI3 and MBPT2 data sets. Moreover the differences of two theoretical  data sets vary smoothly along the isoelectronic sequence. Therefore, these values compiled by the NIST ASD, which have been listed in Table~\ref{table.nist}, seem to be wrong or at least are affected by large errors.

Recently, using the MR-MP  method \citet{Santana.2018.V238.p34} reported calculated excitation energies for L-shell ions of Argon,  including the $n \leq 3$ levels in Ar XI. Compared with the NIST experimental results (excluding seven cases for Ar XI in Table~\ref{table.nist}),  both the present MCDHF/RCI calculations and the MR-MP calculations show good agreements. The average absolute difference with the standard deviation between computed excitation energies and the NIST values for the $n \leq 3$ levels in Ar XI are $-97 \pm 732$ cm$^{-1}$ for MCDHF/RCI, and $-221 \pm 886$ cm$^{-1}$ for MR-MP. High accuracy is achieved in both the MCDHF/RCI and MR-MP calculations.

\subsection{Transition rates}
Wavelengths $\lambda_{ij}$, transition
rates $A_{ji}$, and branching fractions 
(${\rm BF}_{ji} = A_{ji}/ \sum \limits_{k=1}^{j-1} A_{jk}$)
involving all levels considered in the present MCDHF/RCI calculations, as reported in Table~\ref{tab.lev.all}, along with line strength $S_{ji}$ and weighted oscillator strengths $gf_{ji}$, are provided in Table \ref{tab.trans.all}.  
E1 and E2 transition data in both length ($l$) and velocity ($v$) forms are given.  
For E1 and E2 transitions, we provide (last column) the uncertainty estimations of line strengths $S$ adopting the NIST ASD~\citep{Kramida.2018.V.p} terminology (AA $\leq$ 1~\%, A$^{+}$ $\leq$ 2~\%, A $\leq$ 3~\%, B$^{+}$ $\leq$ 7~\%, B $\leq$ 10~\%, C$^{+}$  $\leq$ 18~\%,  C $\leq$ 25~\%,  D$^{+}$ $\leq$ 40~\%, D $\leq$ 50~\%, and E $>$ 50~\% )   
and using the method proposed by~\cite{Kramida.2014.V212.p11}. 
For each E1 transition, the deviation $\delta S$  of line strengths $S_l$ in the length form  and $S_v$  in the velocity form  is defined as 
$\delta S$ = $|S_{v} - S_{l}|$/max($S_{v}$,~$S_{l}$). In various ranges of $S$, the averaged uncertainties $\delta S_{av}$ of $\delta S$ for E1 transitions in Cr XVII  are assessed to 0.5~\% for $S \geq 10^{0}$; 0.7~\% for $10^{0} > S \geq 10^{-1}$; 1.3~\% for $10^{-1} > S \geq 10^{-2}$; 3.3~\% for $10^{-2} > S \geq 10^{-3}$; 6.7~\% for $10^{-3} > S \geq 10^{-4}$, 11~\% for $10^{-4} > S \geq 10^{-5}$, 15~\% for $10^{-5} > S \geq 10^{-6}$, 28~\% for $10^{-6} > S \geq 10^{-7}$. Then,  the largest of $\delta S_{av}$  and $\delta S_{ij}$ is considered to be the uncertainty of each particular transition.

In the spirit of the uncertainty estimation approach~\citep{Kramida.2013.V63.p313,Kramida.2014.V212.p11}, the estimated uncertainties of line strengths $S$ for E2  transitions in Cr XVII  are estimated, as well as those for E1 and E2 transitions in Ar XI, K XII, Ca XIII, Sc XIV, Ti XV, and V~XVI. The estimated uncertainties for all E1 and E2 transitions with BF  $\geq 10^{-5}$ in Ar XI, K XII, Ca XIII, Sc XIV, Ti XV, V~XVI, and Cr XVII, are listed in Table~\ref{tab.trans.all}.

Our MCDHF/RCI radiative lifetimes $\tau_{\rm MCDHF/RCI}^l$ (in s) in the length  form and $\tau_{\rm MCDHF/RCI}^v$ (in s)  in the velocity form, for the lowest 156 (179, 184, 196, 200, 200, 200) states of the $n \leq 3$ configurations in  Ar XI  (K XII, Ca XIII, Sc XIV, Ti XV, V~XVI, Cr XVII), which are calculated by considering all possible E1, E2, M1, and M2 transitions, are provided in Table~\ref{tab.lev.all}. Our MCDHF/RCI radiative lifetimes $\tau_{\rm MCDHF/RCI}^l$ and  $\tau_{\rm MCDHF/RCI}^v$ show good agreement. For example, the average deviation between $\tau_{\rm MCDHF/RCI}^l$ and  $\tau_{\rm MCDHF/RCI}^v$ for all 200 levels in Cr XVII is 1~\%.

\section{Conclusion}
Employing a state-of-the-art method, MCDHF/RCI, the excitation energies and lifetimes of the lowest 156 (179, 184, 196, 200, 200, 200) states of the $n \leq 3$ configurations in  Ar XI  (K XII, Ca XIII, Sc XIV, Ti XV, V~XVI, Cr XVII), have been calculated. Wavelengths, line strengths, transition rates, and oscillator strengths for
the E1, M1, and E2 transitions with BF larger than $10^{-5}$ are also reported. 

Our detailed discussion of excitation energies of these seven O-like ions have highlighted several 
discrepancies in the experimental energies in the NIST ASD. Those levels  compiled by the NIST ASD  have been  listed in Table~\ref{table.nist}. 
The above comparisons clearly show the importance of the present ab initio
calculations to assess the correctness of level and line identifications. Further experimental work is encouraged 
to confirm our suggestions. We believe the present data could serve as benchmarks in future line identifications, and could make important contributions to modeling and diagnosing hot astrophyscial and laboratory plasmas.

\section*{Acknowledgments}
We acknowledge the support from the National Key Research and Development Program of China under Grant No.~2017YFA0403200, the Science Challenge Project of China Academy of Engineering Physics (CAEP) under Grant No. TZ2016005, the National Natural Science Foundation of China (Grant No.~11703004, No.~11674066, No.~11504421, and No.~11734013), the Natural Science Foundation of Hebei Province, China (A2019201300), and the Natural Science Foundation of Educational Department of Hebei Province, China (BJ2018058). This work is also supported by the Fonds de la Recherche Scientifique - (FNRS) and the Fonds Wetenschappelijk Onderzoek - Vlaanderen (FWO) under EOS Project n$^{\rm o}$~O022818F, and by the Swedish research council under contracts 2015-04842 and 2016-04185.  
Kai Wang expresses his gratitude to the support from the visiting researcher program at the Fudan University.


\bibliographystyle{elsarticle-num}
\bibliography{ref.bib}

\onecolumn


\clearpage

\newpage
\section*{Tables}
\setlength{\tabcolsep}{0.5\tabcolsep}
\renewcommand{\arraystretch}{0.7}
\footnotesize 
\begin{longtable}{@{\extracolsep\fill}llllll@{}}
	\caption{The levels in Table~\ref{tab.lev.all} for which our $LSJ$-coupled labels are different from those in the NIST ASD.  \vskip0.5pc \label{table.diff}}
	Z & Key & \multicolumn{2}{c}{MCDHF/RCI} & \multicolumn{2}{c}{NIST}    \\
	\cline{3-4} \cline{5-6}
	&  & Cofiguration   &Term & Cofiguration   &Term  \\
	\hline\\
	\endfirsthead\\
	Z & Key & \multicolumn{2}{c}{MCDHF/RCI} & \multicolumn{2}{c}{NIST}    \\
\cline{3-4} \cline{5-6}
&  & Cofiguration   &Term & Cofiguration   &Term  \\
	\hline\\
	\endhead
19	& 83  &  $2s^{2}\,2p^{3}(^{2}_{1}P)~^{2}P\,3d$    &  $~^{3}P_{2}^{\circ}$  &  $2s^{2}\,2p^{3}(^{2}_{1}P)~^{2}P\,3d$    &  $~^{3}F_{2}^{\circ}$  \\
20	& 65  &  $2s^{2}\,2p^{3}(^{2}_{3}D)~^{2}D\,3d$    &  $~^{1}P_{1}^{\circ}$  &  $2s^{2}\,2p^{3}(^{2}_{3}D)~^{2}D\,3d$    &  $~^{3}D_{1}^{\circ}$  \\
20	& 66  &  $2s^{2}\,2p^{3}(^{2}_{3}D)~^{2}D\,3d$    &  $~^{3}D_{2}^{\circ}$  &  $2s^{2}\,2p^{3}(^{2}_{3}D)~^{2}D\,3d$    &  $~^{3}P_{2}^{\circ}$  \\
20	& 67  &  $2s^{2}\,2p^{3}(^{2}_{3}D)~^{2}D\,3d$    &  $~^{3}D_{1}^{\circ}$  &  $2s^{2}\,2p^{3}(^{2}_{3}D)~^{2}D\,3d$    &  $~^{1}P_{1}^{\circ}$  \\
20	& 83  &  $2s^{2}\,2p^{3}(^{2}_{1}P)~^{2}P\,3d$    &  $~^{3}P_{2}^{\circ}$  &  $2s^{2}\,2p^{3}(^{2}_{1}P)~^{2}P\,3d$    &  $~^{3}F_{2}^{\circ}$  \\
20	& 84  &  $2s^{2}\,2p^{3}(^{2}_{1}P)~^{2}P\,3d$    &  $~^{3}D_{2}^{\circ}$  &  $2s^{2}\,2p^{3}(^{2}_{1}P)~^{2}P\,3d$    &  $~^{1}D_{2}^{\circ}$  \\
20	& 88  &  $2s^{2}\,2p^{3}(^{2}_{1}P)~^{2}P\,3d$    &  $~^{1}D_{2}^{\circ}$  &  $2s^{2}\,2p^{3}(^{2}_{1}P)~^{2}P\,3d$    &  $~^{3}D_{2}^{\circ}$  \\
21	& 84  &  $2s^{2}\,2p^{3}(^{2}_{1}P)~^{2}P\,3d$    &  $~^{3}D_{2}^{\circ}$  &  $2s^{2}\,2p^{3}(^{2}_{1}P)~^{2}P\,3d$    &  $~^{1}D_{2}^{\circ}$  \\
\end{longtable}
\clearpage

\setlength{\tabcolsep}{0.5\tabcolsep}
\renewcommand{\arraystretch}{0.7}
\footnotesize 
\begin{longtable}{@{\extracolsep\fill}lllcccccccccc@{}}
	\caption{Comparison of the present MCDHF/RCI excitation energies (in cm$^{-1}$) with the experimental and theoretical values for the lowest 10 levels of O-like ions with\\ 
$Z= 18 - 24$. NIST: experimental excitation energies compiled in the NIST ASD~\citep{Kramida.2018.V.p}; MCDHF/RCI: the present MCDHF/RCI excitation energies; MCDHF/RCI2: the \\
MCDHF/RCI excitation energies from \citet{Rynkun.2013.V557.p136}; MBPT: the MBPT results from \citet{Gu.2005.V89.p267}; SS: the SUPERSTRUCTURE results from \citet{Landi.2005.V434.p365}.   Differences \\ ($\Delta E_x= E_x- E_{\rm NIST}$) 
in cm$^{-1}$ from different theoretical results with the NIST values are also included. \vskip0.5pc
	\label{table.n2}}
	Z & Key & Level & &  &$E$ &  & & &$\Delta E$ & &    \\
	\cline{5-8} \cline{9-12}
	&  &  &   NIST & MCDHF/RCI  & MCDHF/RCI2 & MBPT & SS & MCDHF/RCI & MCDHF/RCI2 & MBPT &  SS  \\
	\hline\\
	\endfirsthead\\
	\caption[]{($continued$)}
	Z & Key & Level & &  &$E$ &  & & &$\Delta E$ & &    \\
\cline{5-8} \cline{9-12}
&  &  &   NIST & MCDHF/RCI  & MCDHF/RCI2 & MBPT & SS & MCDHF/RCI & MCDHF/RCI2 & MBPT &  SS  \\
	\hline\\
	\endhead
	18 &   1    & $2s^{2}\,2p^{4}(^{3}_{2}P)~^{3}P_{2}$      & 0         & 0            &   0           & 0       & 0        &      0     &       0         & 0       & 0        \\
	18 &   2    & $2s^{2}\,2p^{4}(^{3}_{2}P)~^{3}P_{1}$      &  14453.4  & 14436        &   14457       & 14453   & 14878    &    -17.4   &     3.6         & -0.4    & 424.6   \\
	18 &   3    & $2s^{2}\,2p^{4}(^{3}_{2}P)~^{3}P_{0}$      & 18322     & 18458        &   18308       & 18310   & 18810    &     136    &     -14         & -12     & 488     \\
	18 &   4    & $2s^{2}\,2p^{4}(^{1}_{2}D)~^{1}D_{2}$      &  71833.4  & 72022        &   71967       & 71853   & 75409    &     188.6  &   133.6         & 19.6    & 3575.6  \\
	18 &   5    & $2s^{2}\,2p^{4}(^{1}_{0}S)~^{1}S_{0}$      & 148537.1  & 148721       &   148833      & 149258  & 150351   &     183.9  &   295.9         & 720.9   & 1813.9  \\
	18 &   6    & $2s~^{2}S\,2p^{5}~^{3}P_{2}^{\circ}$       & 529604    & 529442       &   529877      & 529020  & 540055   &    -162    &     273         & -584    & 10451   \\
	18 &   7    & $2s~^{2}S\,2p^{5}~^{3}P_{1}^{\circ}$       & 541973    & 541776       &   542179      & 541328  & 552681   &    -197    &     206         & -645    & 10708   \\
	18 &   8    & $2s~^{2}S\,2p^{5}~^{3}P_{0}^{\circ}$       & 548984    & 548842       &   549224      & 548355  & 559875   &    -142    &     240         & -629    & 10891   \\
\end{longtable}
\clearpage

\setlength{\tabcolsep}{0.5\tabcolsep}
\renewcommand{\arraystretch}{0.7}
\footnotesize 
\begin{longtable}{@{\extracolsep\fill}lllcccccccccc@{}}
	\caption{Excitation energies (in cm$^{-1}$)  of O-like Cr XVII. MCDHF/RCI: the present MCDHF/RCI results; MCDHF/RCI3 -- the MCDHF/RCI results from \citet{Wang.2017.V229.p37}; \\ MBPT2: the MBPT results from \citet{Wang.2017.V229.p37}; CI: theoretical  results from \citet{Bogdanovich.2008.V94.p623}. NIST: experimental excitation energies compiled \\ in the NIST ASD~\citep{Kramida.2018.V.p}. $\Delta E$ -- differences  (in cm$^{-1}$)  between different sources (see above) and the MCDHF/RCI values are also included.  \vskip0.5pc
		\label{table.Cr}}
	&  & MCDHF/RCI &    \multicolumn{2}{c}{MCDHF/RCI3}&  \multicolumn{2}{c}{MBPT2} &\multicolumn{2}{c}{CI} & \multicolumn{2}{c}{NIST} \\
	\cline{3-3} \cline{4-5} \cline{6-7}\cline{8-9} \cline{10-11}
	Key  & Level &   $E$ & $E$  & $\Delta E$ & $E$ & $\Delta E$ & $E$ & $\Delta E$ & $E$ & $\Delta E$  \\
	\hline\\
	\endfirsthead\\
	\caption[]{($continued$)}
	&  & MCDHF/RCI &    \multicolumn{2}{c}{MCDHF/RCI3}&  \multicolumn{2}{c}{MBPT2} &\multicolumn{2}{c}{CI} & \multicolumn{2}{c}{NIST} \\
\cline{3-3} \cline{4-5} \cline{6-7}\cline{8-9} \cline{10-11}
Key  & Level &   $E$ & $E$  & $\Delta E$ & $E$ & $\Delta E$ & $E$ & $\Delta E$ & $E$ & $\Delta E$  \\
	\hline\\
	\endhead
1    &  $2s^{2}\,2p^{4}(^{3}_{2}P)~^{3}P_{2}$                             &   0          &    0        & 0        & 0        &  0     &  0         &  0        &  0          & 0            \\
2    &  $2s^{2}\,2p^{4}(^{3}_{2}P)~^{3}P_{0}$                             &   58142      &    58159    & 17       & 58117    &  -25   &  58315     &  173      &  57680      & -462         \\
3    &  $2s^{2}\,2p^{4}(^{3}_{2}P)~^{3}P_{1}$                             &   60376      &    60380    & 4        & 60456    &  80    &  60551     &  175      &  60376      & 0            \\
4    &  $2s^{2}\,2p^{4}(^{1}_{2}D)~^{1}D_{2}$                             &   135161     &    135132   & -29      & 134970   &  -191  &  134872    &  -289     &  134980     & -181         \\
5    &  $2s^{2}\,2p^{4}(^{1}_{0}S)~^{1}S_{0}$                             &   263198     &    263290   & 92       & 263008   &  -190  &  262826    &  -372     &  262970     & -228         \\
6    &  $2s~^{2}S\,2p^{5}~^{3}P_{2}^{\circ}$                              &   812691     &    813320   & 629      & 812560   &  -131  &  812005    &  -686     &  813180     & 489          \\
7    &  $2s~^{2}S\,2p^{5}~^{3}P_{1}^{\circ}$                              &   857169     &    857798   & 629      & 857071   &  -98   &  856388    &  -781     &  857690     & 521          \\
8    &  $2s~^{2}S\,2p^{5}~^{3}P_{0}^{\circ}$                              &   887059     &    887702   & 643      & 886984   &  -75   &  886318    &  -741     &  887540     & 481          \\
\end{longtable}
\clearpage

\setlength{\tabcolsep}{0.5\tabcolsep}
\renewcommand{\arraystretch}{0.7}
\footnotesize 
\begin{longtable}{@{\extracolsep\fill}lllllrc@{}}
	\caption{Excitation energies (in cm$^{-1}$)  for levels for which which the NIST values differ with the present MCDHF/RCI values by more than 3000 cm$^{-1}$ (about 0.1~\%). \\ $E_{\rm MCDHF/RCI}$: the present MCDHF/RCI values; $\Delta E_{\rm NIST}$: experimental excitation energies compiled  in the NIST ASD~\citep{Kramida.2018.V.p};
		$\Delta E  = E_{\rm MCDHF/RCI} -E_{\rm MCDHF}$; \\
Reference: reference sources for the NIST values.		\vskip0.5pc
		\label{table.nist}}
	Z  & Key & Level &  $E_{\rm{MCDHF/RCI}}$ & $E_{\rm{NIST}}$  &$\Delta E$ & Reference \\
	\hline\\
	\endfirsthead\\
	\caption[]{($continued$)}
	Z  & Key & Level &  $E_{\rm{MCDHF/RCI}}$ & $E_{\rm{NIST}}$  &$\Delta E$ & Reference \\
	\hline\\
	\endhead
18  &    26   &   $2s^{2}\,2p^{3}(^{2}_{1}P)~^{2}P\,3s~^{1}P_{1}^{\circ}$    &  2660649	  &     2649800	     &     -10849    &  \cite{Saloman.2010.V39.p33101}            \\
18  &    31   &   $2s^{2}\,2p^{3}(^{2}_{3}D)~^{2}D\,3p~^{3}F_{2}$            &  2724663	  &     2728790	     &     4127      &  \cite{Saloman.2010.V39.p33101}            \\
18  &    61   &   $2s^{2}\,2p^{3}(^{2}_{3}D)~^{2}D\,3d~^{3}G_{3}^{\circ}$    &  2896390	  &     2900300	     &     3910      &  \cite{Saloman.2010.V39.p33101}            \\
18  &    72   &   $2s^{2}\,2p^{3}(^{2}_{3}D)~^{2}D\,3d~^{3}P_{1}^{\circ}$    &  2925253	  &     2932610	     &     7357      &  \cite{Saloman.2010.V39.p33101}            \\
18  &    79   &   $2s^{2}\,2p^{3}(^{2}_{1}P)~^{2}P\,3d~^{3}F_{3}^{\circ}$    &  2954139	  &     2945040	     &     -9099     &  \cite{Saloman.2010.V39.p33101}            \\
18  &    81   &   $2s^{2}\,2p^{3}(^{2}_{1}P)~^{2}P\,3d~^{3}P_{1}^{\circ}$    &  2957524	  &     2968000	     &     10476     &  \cite{Saloman.2010.V39.p33101}            \\
18  &    85   &   $2s^{2}\,2p^{3}(^{2}_{1}P)~^{2}P\,3d~^{3}D_{1}^{\circ}$    &  2971222	  &     2987530	     &     16308     &  \cite{Saloman.2010.V39.p33101}            \\
19  &    67   &   $2s^{2}\,2p^{3}(^{2}_{3}D)~^{2}D\,3d~^{3}D_{1}^{\circ}$    &  3361421	  &     3358100      &     -3321     &  \cite{Fawcett.1975.V170.p185}              \\
\end{longtable}

\normalsize

\clearpage

\newpage

\TableExplanation

\bigskip
\renewcommand{\arraystretch}{1.0}

\section*{Table ~1.\label{tbl1te1}Excitation energies in cm$^{-1}$ and radiative lifetimes in s for the lowest 156 (179, 184, 196, 200, 200, 200) states of the $n \leq 3$ configurations in  Ar~XI  (K~XII, Ca~XIII, Sc~XIV, Ti~XV, V~XVI, Cr~XVII).
}
\begin{tabular*}{0.95\textwidth}{@{}@{\extracolsep{\fill}}lp{5.5in}@{}}
	Key             &A number assigned to each level \\
	Level  &The $LSJ$-coupled label for each level \\
	$E_{\rm NIST}$  & observed values compiled in  the NIST ASD  \citep{Kramida.2018.V.p}\\
	$E_{\rm MCDHF/RCI}$ & our MCDHF/RCI excitation energies\\
	$\tau_{\rm MCDHF/RCI}^l$ & our MCDHF/RCI lifetimes in length form\\
	$\tau_{\rm MCDHF/RCI}^v$ & our MCDHF/RCI lifetimes  in velocity form\\
	$LS$-composition & the $LS$ eigenvector compositions \\
\end{tabular*}

\section*{Table ~2.\label{tbl2te1} Transition wavelengths $\lambda$ (in {\AA}), transition rates $A$ (in s$^{-1}$), weighted oscillator strengths $gf$, and line strengths $S$ (in a.u.) between  the lowest 156 (179, 184, 196, 200, 200, 200) states of the $n \leq 3$ configurations in  Ar~XI  (K~XII, Ca~XIII, Sc~XIV, ~~{Ti~XV}, V~XVI, Cr~XVII) listed in Table~\ref{tab.lev.all}. 
	Transitions with the branching fraction (BF) $\geq 10^{-5}$ are presented.}
\begin{tabular*}{0.95\textwidth}{@{}@{\extracolsep{\fill}}lp{5.5in}@{}}
	Z             &The nuclear charge number \\	
	$i$         &The lower level of a transition \\
	$j$         &The upper level of a transition \\
	$\lambda$  &Wavelength, in \AA. \\
	Type        &Transition type (E1, M1, E2, M2). \\
	BF         &The branching fraction from the upper level. \\
	$A^l$ & Transition rate (in s$^{-1}$) in the length ($l$)  form\\ 
	$gf^l$ & Weighted oscillator strength (dimensionless)  in the length ($l$)  form\\
	$S^l$ & Line strength (in a.u.)   in the length ($l$)  form\\
	$A^v$ & Transition rate (in s$^{-1}$) in the  velocity ($v$)  form\\ 
	$gf^v$ & Weighted oscillator strength (dimensionless)  in the  velocity ($v$)  form\\
	$S^v$&Line strength (in a.u.)   in the  velocity ($v$)  form\\
	Acc. & the estimated  accuracies of the $S$ values using the terminologies (AA $\leq$ 1~\%, A$^{+}$ $\leq$ 2~\%, A $\leq$ 3~\%, B$^{+}$ $\leq$ 7~\%, B $\leq$ 10~\%, C$^{+}$  $\leq$ 18~\%,  C $\leq$ 25~\%,  D$^{+}$ $\leq$ 40~\%, D $\leq$ 50~\%, and E $>$ 50~\%) of the NIST ASD  \citep{Kramida.2018.V.p}. 
\end{tabular*}

\renewcommand{\arraystretch}{1.0}
\bigskip

\datatables 


\setlength{\LTleft}{0pt} \setlength{\LTright}{0pt}

\linespread{1}
\tiny
\setlength{\tabcolsep}{5pt}
\begin{longtable}{llllllll}
	\caption{Excitation energies in cm$^{-1}$ and radiative lifetimes in s for the lowest 156 (179, 184, 196, 200, 200, 200) states of the $n \leq 3$ configurations in  Ar~XI  (K~XII, Ca~XIII, Sc~XIV, Ti~XV, V~XVI, Cr~XVII). 
		\label{tab.lev.all}}\\
	Z & Key & Level & $E_{\rm NIST}$ & $E_{\rm MCDHF/RCI}$ & $\tau_{\rm MCDHF/RCI}^l$ & $\tau_{\rm MCDHF/RCI}^v$&$LS$-composition\\
	\midrule
	\endfirsthead
	\midrule
	Z & Key & Level & $E_{\rm NIST}$ & $E_{\rm MCDHF/RCI}$ & $\tau_{\rm MCDHF/RCI}^l$ & $\tau_{\rm MCDHF/RCI}^v$&$LS$-composition\\
	\midrule
	\endhead
	\endfoot
	\bottomrule
	\endlastfoot
18 & 1   & $2s^{2}\,2p^{4}(^{3}_{2}P)~^{3}P_{2}$                                            &    0       &0       &  ...       & ...       &  0.98                                                                                                                                  \\ 
18 & 2   & $2s^{2}\,2p^{4}(^{3}_{2}P)~^{3}P_{1}$                                            & 14453.4    &14436   &  1.504E-02 & 1.504E-02 &  0.99                                                                                                                                  \\ 
18 & 3   & $2s^{2}\,2p^{4}(^{3}_{2}P)~^{3}P_{0}$                                            & 18322      &18458   &  2.908E-01 & 2.908E-01 &  0.97                                                                                                                                  \\ 
18 & 4   & $2s^{2}\,2p^{4}(^{1}_{2}D)~^{1}D_{2}$                                            & 71833.4    &72022   &  3.646E-03 & 3.646E-03 &  0.98                                                                                                                                  \\ 
18 & 5   & $2s^{2}\,2p^{4}(^{1}_{0}S)~^{1}S_{0}$                                            & 148537.1   &148721  &  3.953E-04 & 3.953E-04 &  0.95 + 0.03~$2p^{6}~^{1}S$                                                                                                            \\ 
18 & 6   & $2s~^{2}S\,2p^{5}~^{3}P_{2}^{\circ}$                                             & 529604     &529442  &  4.806E-11 & 4.792E-11 &  0.99                                                                                                                                  \\ 
18 & 7   & $2s~^{2}S\,2p^{5}~^{3}P_{1}^{\circ}$                                             & 541973     &541776  &  4.608E-11 & 4.611E-11 &  0.99                                                                                                                                  \\
\end{longtable}

\clearpage
\linespread{1}
\tiny
\setlength{\tabcolsep}{5pt}
\begin{longtable}{llccccccccccccc}
	\caption{Transition wavelengths $\lambda$ (in {\AA}), transition rates $A$ (in s$^{-1}$), weighted oscillator strengths $gf$, and line strengths $S$ (in a.u.) 
		between  the lowest 156 (179, 184, 196, 200, 200, 200) states of the $n \leq 3$ configurations in  Ar~XI  (K~XII, Ca~XIII, Sc~XIV, Ti~XV, \\V~XVI, Cr~XVII) listed in Table~\ref{tab.lev.all}. 
	   Transitions with the branching fraction (BF) $\geq 10^{-5}$ are presented. \label{tab.trans.all} }\\
	\toprule
$Z$ & $i$ & $j$ &    
$\lambda$ & Type  &BF & $A^l$ & $gf^l$ & $S^l$ & $A^v$ & $gf^v$ & $S^v$&  Acc.  \\	\midrule
	\endfirsthead
\midrule
$Z$ & $i$ & $j$ &    
$\lambda$ & Type  &BF & $A^l$ & $gf^l$ & $S^l$ & $A^v$ & $gf^v$ & $S^v$&  Acc.  \\		
	\midrule
	\endhead
	\midrule
	\endfoot
	\bottomrule
	\endlastfoot
18&    1&    2&  6.927323E+03&  M1&  1.000E+00&  6.649E+01&  1.435E-06&  2.458E+00&           &           &           &       \\      
18&    1&    3&  5.417764E+03&  E2&  6.328E-04&  2.176E-03&  9.577E-12&  9.070E-03&  1.995E-03&  8.778E-12&  8.314E-03&   C   \\      
18&    2&    3&  2.486195E+04&  M1&  9.994E-01&  3.437E+00&  3.185E-07&  1.958E+00&           &           &           &       \\      
18&    1&    4&  1.388462E+03&  M1&  8.505E-01&  2.333E+02&  3.371E-07&  1.157E-01&           &           &           &       \\      
18&    1&    4&  1.388462E+03&  E2&  3.281E-04&  8.999E-02&  1.300E-10&  2.073E-03&  8.496E-02&  1.228E-10&  1.957E-03&   C   \\      
18&    2&    4&  1.736517E+03&  M1&  1.492E-01&  4.092E+01&  9.249E-08&  3.972E-02&           &           &           &       \\      
18&    2&    4&  1.736517E+03&  E2&  1.661E-05&  4.556E-03&  1.030E-11&  3.212E-04&  4.347E-03&  9.825E-12&  3.064E-04&   C   \\      
18&    1&    5&  6.723995E+02&  E2&  3.276E-04&  8.288E-01&  5.618E-11&  1.017E-04&  8.289E-01&  5.618E-11&  1.017E-04&   C   \\
\end{longtable}

\end{document}